\documentclass[aps,twocolumn,floats,tightenlines,balancelastpage,floatfix,nofootinbib,showpacs,preprintnumbers]{revtex4}
\input epsf
\usepackage{amsmath,amssymb}
\usepackage{graphicx}

\newcommand{\mnras}{Mon. Not. R. Astron. Soc.}

\newcommand{\lsim}{\mathrel{\hbox{\rlap{\lower.55ex\hbox{$\sim$}} \kern-.3em \raise.4ex \hbox{$<$}}}}
\newcommand{\gsim}{\mathrel{\hbox{\rlap{\lower.55ex\hbox{$\sim$}} \kern-.3em \raise.4ex \hbox{$>$}}}}
\begin{document}
\title{Axion constraints in non-standard thermal histories}
\author{Daniel Grin, Tristan L. Smith, and Marc Kamionkowski} 
\affiliation{California Institute of Technology, Mail Code 130-33, Pasadena, CA 91125}
\date{\today}
\begin{abstract} 
It is usually assumed that dark matter is produced during the radiation dominated era. There is, however, no direct evidence for radiation domination prior to big-bang nucleosynthesis. Two non-standard thermal histories are considered. In one, the low-temperature-reheating scenario, radiation domination begins as late as $\sim 1$ MeV, and is preceded by significant entropy generation. Thermal axion relic abundances are then suppressed, and cosmological limits to axions are loosened. For reheating temperatures $T_{\rm rh}\lsim 35~{\rm MeV}$, the large-scale structure limit to the axion mass is lifted. The remaining constraint from the total density of matter is significantly relaxed. Constraints are also relaxed for higher reheating temperatures. In a kination scenario, a more modest change to  cosmological axion constraints is obtained. Future possible constraints to axions and low-temperature reheating from the helium abundance and next-generation large-scale-structure surveys are discussed.
\end{abstract}

\pacs{14.80.Mz,98.80.-k,95.35.+d}
\maketitle
\section{Introduction}
The Peccei-Quinn (PQ) solution to the strong-CP problem yields the axion, a dark-matter candidate \cite{crewther,baluni,pq}. If the axion mass $m_{\rm a}\gsim 10^{-2}~{\rm eV}$, axions will be produced thermally, with cosmological abundance \begin{equation}\Omega_{{\rm a}}h^{2}=\frac{m_{{\rm a}}}{130~{\rm eV}}\left(\frac{10}{g_{*_{\rm S},\rm F}}\right)\label{thermalfreeze},\end{equation} where $g_{*_{\rm S},{\rm F}}$ is the effective number of relativistic degrees of freedom when axions freeze out \cite{changchoi,notinvisible,kt,murayama}. Axions with masses in the $\sim {\rm eV}$ range would contribute to the total density in roughly equal proportion to baryons.

Axions in the $\sim{\rm eV}$ mass range are relativistic when they decouple at $T_{{\rm F}}= 30-50~{\rm MeV}$ \cite{changchoi}. Free streaming then erases density perturbations, suppressing the matter power spectrum on scales smaller than the axion free-streaming length \cite{hu1,hu2,kt,haneutrino5}. Light axions would also contribute to the early integrated Sachs-Wolfe (ISW) effect \cite{isw}. Data from large-scale structure (LSS) surveys and cosmic microwave-background (CMB) observations have been used to impose the constraint $m_{\rm a}\lsim1~{\rm eV}$ to light hadronic axions \cite{hanraffelt,raffeltwong,slosar}. These arguments apply to any particle relativistic at matter-radiation equality or cosmic microwave background (CMB) decoupling, thus imposing the similar constraint
$\sum_{i} m_{\nu,i}\lsim 1~{\rm eV}$ to the sum of neutrino masses \cite{crotty,tegmarksdss,barger,seljakneutrino,haneutrino5,fogli2,kristiansen,fukugitawmap3,spergelwmap3,elgaroy,pierpaoli}.

These constraints rely on abundances computed assuming that radiation domination began earlier than the chemical freeze-out of light relics. There is, however, no direct evidence for radiation domination prior to big-bang nucleosynthesis (BBN) \cite{kamion}. The transition to radiation domination may be more gradual than typically assumed. In such a modified thermal history, two effects may cause relic abundances to change. First, the Hubble expansion rate scales differently with temperature $T$ until radiation domination begins, leading to a different freeze-out temperature. Second, entropy may be generated, suppressing relic abundances.  

The universe could have reheated to a temperature as low as $1~{\rm MeV}$, with standard radiation domination beginning thereafter \cite{kawasaki,kawasaki2,ichikawa3,giudice2,hannestadreheat,kolbnotario}. This low-temperature reheating (LTR) scenario may be modeled simply through the entropy-generating decay of a massive particle $\phi$ into radiation, with fixed rate $\Gamma_{\phi}$ and initial value $H_{\rm I}$ of the Hubble parameter. The scalar $\phi$ may be the inflaton, oscillating as inflation ends and decaying into standard-model particles, or it might be a secondary scalar, produced during preheating \cite{dolgovlinde,afwise,kofman1,kofman3,shtanovreheat}. This decay softens the scaling of temperature $T$ with cosmological scale factor $a$, increasing the Hubble parameter $H(T)$ and leading to earlier freeze-out for certain relics. Entropy generation then highly suppresses these relic abundances.

Kination models offer another alternative to the standard thermal history, invoking a period of scalar-field kinetic-energy dominance \cite{kination}, but no entropy production. Without entropy generation, abundances change more modestly.

Past work has shown that cosmological constraints to neutrinos, weakly interacting massive particles, and non-thermally produced axions are relaxed in LTR \cite{giudice1,giudice2,yaguna}. Non-thermally produced axions ($m_{\rm a}\lsim 10^{-2}~{\rm eV}$) would be produced through coherent oscillations of the PQ pseudoscalar \cite{notinvisible, kt,torig}. In this paper, we obtain constraints to thermally-produced hadronic axions in the kination and LTR scenarios. While kination modestly loosens limits, LTR dramatically changes the cosmologically allowed range of axion masses.

We begin by reviewing these modified thermal histories and calculating axion relic abundances. We then generalize cosmological constraints to axions, allowing for low-temperature reheating and kination. For reheating temperatures $T_{\rm rh} \lsim 35~{\rm MeV}$, LSS/CMB limits to the axion mass are lifted; constraints are also relaxed for higher $T_{\rm rh}$. Constraints from the total matter density are also relaxed, but not completely lifted. For $T_{\rm rh}\simeq 10~{\rm MeV}$, the new constraint is $m_{\rm a}\lsim 1.4~{\rm keV}$, while for $T_{\rm rh}\simeq 35~{\rm MeV}$, we find that $m_{\rm a}\lsim 43~{\rm eV}$. If $T_{\rm rh}\gsim170~{\rm MeV}$, standard results are recovered. After estimating the ability of
future large-scale-structure surveys to further constrain axion masses for a variety of reheating temperatures, we derive modestly relaxed constraints to axions in the kination scenario. We conclude by considering future possible constraints to the relativistic energy density of axions in a low-temperature reheating model.
\section{Two non-standard thermal histories: Low-temperature reheating and kination}
\label{ltrsec}
We now review the low-temperature reheating (LTR) scenario. We consider the coupled evolution of unstable massive particles $\phi$, which drive reheating, and radiation $R$, both in kinetic equilibrium. The relevant distribution functions obey a Boltzmann equation with a decay term, and may be integrated to yield \cite{chung,giudice1,giudice2}:
\begin{equation}
\frac{1}{a^{3}}\frac{d\left(\rho_{\phi}a^{3}\right)}{dt}=-\Gamma_{\phi}\rho_{\phi}~~
\frac{1}{a^{4}}\frac{d\left(\rho_{R}a^{4}\right)}{dt}=\Gamma_{\phi}\rho_{\phi},\label{axabev}\end{equation} where $\rho_{\phi}$ and $\rho_{R}$ denote the energy densities in the scalar field and radiation, respectively, $\Gamma_{\phi}$ is the decay rate of the scalar to radiation, and $a$ is the cosmological scale factor. The evolution of the scale factor is given by the Friedmann equation, which is $H^{2}=[8\pi/(3M_{{\rm pl}}^{2})]\left(\rho_{\phi}+\rho_{R}\right)$ well before matter or vacuum-energy domination. The reheating temperature $T_{\rm rh}$ is defined by \cite{giudice1,chung,kt}
\begin{equation}
\Gamma_{\phi}\equiv \sqrt{\frac{4\pi^{3}g_{*,\rm rh}}{45}}\frac{T_{{\rm rh}}^{2}}{M_{{\rm pl}}},
\end{equation}where $M_{{\rm pl}}$ is the Planck mass and $g_{*,\rm rh}$ is the effective number of relativistic degrees of freedom when $T=T_{\rm rh}$. In our calculation of the expansion history in LTR, we use $g_{*}$ calculated using the methods of Refs.~\cite{kt,gongel}, as tabulated for use in the DarkSUSY package \cite{darksusy}. We neglect the axionic contribution to $g_{*}$ for simplicity and assume $3$ massless neutrinos. The resulting $\sim 10\%$ error in $g_{*}$ leads to a comparable fractional error in the resulting axion relic abundance, and is thus negligible at our desired level of accuracy.

\begin{figure}[phtb]
\includegraphics[height=3.2in]{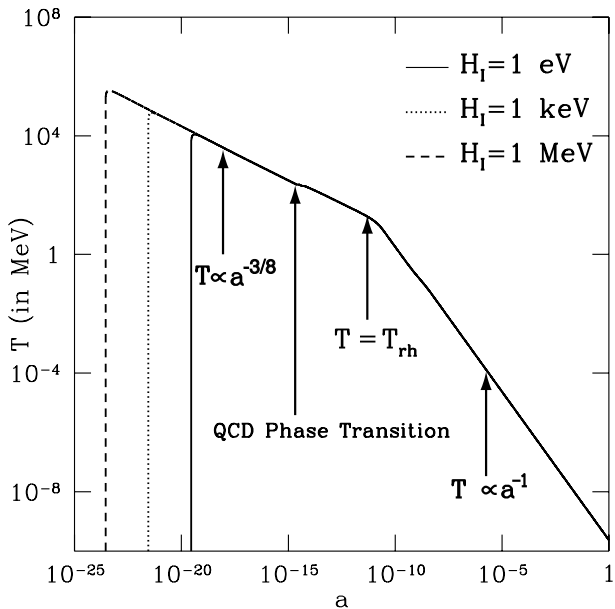}
\caption{This plot shows the evolution of temperature with scale factor in a low-temperature reheating (LTR) cosmology with $T_{\rm rh}=20~{\rm MeV}$ and $3$ different initial values for the Hubble parameter $H_{{\rm I}}$. After a rapid rise due to $\phi$ decay, $T\propto a^{-3/8}$ until $T\sim T_{\rm rh}$, after which radiation domination begins, and $T \propto a^{-1}$. The small bump near $T\simeq 200~{\rm MeV}$ results from a jump in $g_{*}$ near the QCD phase transition.}
\label{figureeva}
\end{figure}

\begin{figure}[phtb]
\includegraphics[width=3.3in]{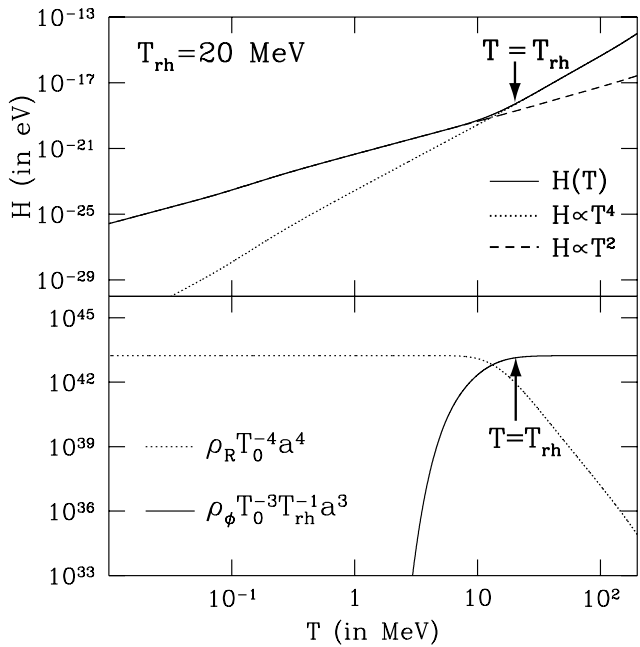}
\caption{The top panel shows the Hubble parameter as a function of temperature in an LTR cosmology with $T_{\rm rh}=20~{\rm MeV}$ and $H_{\rm I}=1~{\rm MeV}$. Initially $H\propto T^{4} g_{*}\left(T\right)$, but at temperatures cooler than $T\sim T_{\rm rh}$, $H\propto T^{2} \sqrt{g_{*}\left(T\right)}$.The second panel shows the comoving radiation energy density $R=\rho_{R}T_ {0}^{-4}a^{4}$ and scalar energy density $\Phi=\rho_{\phi}T_{0}^{-3}{T}_{\rm rh}^{-1}a^{3}$ as a function of temperature. At $T\sim T_{\rm rh}$, $R$ flattens out to a constant and $\Phi$ drops off to zero, indicating the conclusion of reheating.}
\label{figureevb}
\end{figure}

We use dimensionless comoving densities \cite{giudice1,chung}:
\begin{equation}
\Phi\equiv\rho_{\phi}T_{{\rm rh}}^{-1}T_{\rm 0}^{-3}a^{3},~ R\equiv \rho_{R}a^{4}T_{0}^{-4}\label{newvar},\end{equation} where $T_{0}$ is the temperature today. At the beginning of reheating, $\phi$ dominates the energy density and radiation is negligible. Thus, as initial conditions, we use $\rho_{R}=0$ and $\rho_{\phi}=\left[3/\left(8\pi\right)\right]M_{{\rm pl}}^{2}H_{{\rm I}}^{2}$, where $H_{\rm I}$
is the initial value of the Hubble parameter. The two physical free parameters in this model are $T_{\rm rh}$ and $H_{\rm I}$. The temperature is related to the radiation energy density by \cite{kt} \begin{equation}
T=\left[\frac{30}{\pi^{2}g_{*}(T)}\right]^{1/4}\rho_{R}^{1/4}.\label{tr}\end{equation} We numerically integrate Eqs.~(\ref{axabev}) to obtain the dependence of $T$ on $a$, and the results are shown in Fig.~ \ref{figureeva}. As the scalar begins to decay, the temperature rises sharply to a maximum at $a_{\rm m}=\left(8/3\right)^{2/5}a_{\rm I}$, where $a_{\rm I}$ is the initial value of the scale factor, and then falls as $T\propto a^{-3/8}$. This shallow scaling of temperature with scale factor results from the continual dumping of scalar-field energy into radiation, and yields an unusually \textit{steep} dependence of scale factor on temperature. As shown in Figs.~\ref{figureeva} and \ref{figureevb}, when the comoving radiation energy density $R$ overtakes $\Phi$ near $T\sim T_{\rm rh}$, the epoch of radiation domination begins, with the usual $T\propto a^{-1}$ scaling.

Well before reheating concludes, $\Phi$ is constant and $\rho_{\rm R}\ll \rho_{\phi}$. If $a_{\rm m}\ll a<a\left(T_{\rm rh}\right)$, an approximate solution of Eqs.~(\ref{axabev}) for $T(a)$ is then \cite{giudice1,chung}
\begin{eqnarray}\nonumber
T\simeq T_{\rm max} \left(\frac{a}{a_{\rm m}}\right)^{-3/8} \left[\frac{g_{*}\left(T_{m}\right)}{g_{*}\left(T\right)}\right]^{1/4} \\
T_{\rm max}=4.2~{\rm GeV}~\left[\frac{10g_{*,\rm rh}}{g_{*}^{2}\left(T_{\rm m}\right)}\right]^{1/8}
H_{{\rm I}, {\rm eV}}^{1/4}T_{\rm rh, MeV}^{1/2}\label{tsfac},
\end{eqnarray} 
where $g_{*,\rm rh}=g_{*}\left(T_{\rm rh}\right)$ and $T_{\rm max}$ is the maximum temperature obtained (see Fig.~\ref{figureeva}). Here $T_{\rm rh, \rm MeV}$ and $H_{\rm I, \rm eV}$ are the reheating temperature and initial value of the Hubble parameter, in units of ${\rm MeV}$ and ${\rm eV}$, respectively.  

During reheating, the Hubble parameter is given by \cite{giudice1,giudice2}\begin{equation}
H\simeq\left[\frac{5\pi^{3}g_{*}^{2}\left(T\right)}{9g_{*,\rm rh}}\right]^{1/2}\left(\frac{T}{T_{\rm rh}}\right)^{2}\frac{T^{2}}{M_{\rm pl}}.\label{hubblere}
\end{equation} 
At a given temperature, the universe thus expands faster during reheating than it would during radiation domination, and the equilibrium condition $\Gamma\equiv n\left\langle\sigma v\right\rangle\gsim H$ is harder to establish and maintain. Relics with freeze-out temperature $T_{\rm F}\geq T_{\rm max}$ will thus have highly suppressed abundances because they never come into chemical equilibrium. Relics with $T_{\rm rh}\lsim T_{\rm F}\lsim T_{\rm max}$ come into chemical equilibrium, but then freeze out before reheating completes. Their abundances are then reduced by entropy production during reheating. In either case, species with $T_{\rm F}\gsim T_{\rm rh}$ have highly suppressed relic abundances. 

Less radical changes to abundances follow in kination scenarios. During epochs dominated by the kinetic energy of a scalar field, the energy density $\rho$ scales according to $\rho\propto a^{-6}$ \cite{kination}. Thus $H\left(T\right)\propto T^{3}$ or $H\simeq H_{\rm rad}\left(T\right)\left(T/T_{\rm kin}\right)$, where $H_{\rm rad}\left(T\right)$ is the standard radiation-dominated $H\left(T\right)$ and $T_{\rm kin}$ denotes the transition temperature from kination to radiation domination. Kination yields relic freeze-out temperatures somewhere between the standard and LTR values. There is, however, no entropy generation during kination, leading to a less dramatic change in relic abundances.  Note that these conclusions are rather general, as we have not relied on any detailed properties of the kination model, but only on the scaling $H\left(T\right)\propto T^{3}$ \cite{kamion}.

\section{Axion production in non-standard thermal histories}
\label{production}
Axions with $m_{{\rm a}}\gsim 10^{-2}~{\rm eV}$ are thermally produced in the standard radiation-dominated cosmology. We now show that in LTR models, these axions have suppressed abundances. We consider standard hadronic axions, which do not couple to standard-model quarks and leptons at tree level but do have higher-order couplings to pions and photons \cite{kim79,svz80}. We do not consider flaton models, or other scenarios in which PQ symmetry breaking is related to supersymmetry breaking \cite{lyth1,lyth2,lyth3,ss1,ss2}. For temperatures $T\lsim 150~{\rm MeV}$, the dominant channels for axion production are $\pi^{+}+\pi^{-}\to  a+\pi^{0}$, $\pi^{+}+\pi^{0}\to \pi^{+}+ a$, and $\pi^{-}+\pi^{0}\to a+\pi^{-}$. Axion scattering rates are suppressed relative to particle-number-changing interactions by factors of $T^{2}/f_{\rm PQ}^{2}$ and thus decouple very early. Thus, axions stay in kinetic equilibrium because of $\pi+\pi\leftrightarrow a+\pi$, and kinetically decouple when they chemically freeze out. 

Nucleonic channels are negligible at these temperatures. If pions are in chemical equilibrium and Bose enhancement can be neglected, the axion production rate is \cite{hanraffelt,changchoi,kt,khlopovonprod}
\begin{eqnarray}
\nonumber
\Gamma=\frac{3\zeta(3)T^{5}C_{\rm a \pi}^{2}}{1024\pi^{7}f_{\rm a}^{2}f_{\pi}^{2}}\int dx_{1} dx_{2} \frac{x_{1}^{2}x_{2}^{2}}{y_{1}y_{2}}f(y_{1})f(y_{2})\\ \nonumber \times
\int_{-1}^{1}d\mu \frac{\left[s-m_{\pi}^{2}\right]^{3}\left[5s-2m_{\pi}^{2}\right]}{s^{2}T^{4}},\\ C_{{\rm a} \pi}=\frac{1-r}{3\left(1+r\right)}~~~
s=2\left[m_{\pi}^{2}+T^{2}\left(y_{1}y_{2}-x_{1}x_{2}\mu\right)\right],
\label{pionrate}
\end{eqnarray}
where $x_{i}=p_{i}/T$ is the dimensionless pion momenta, $y_{i}=\sqrt{x_{i}^{2}+m_{\pi}^{2}/T^{2}}$ is the dimensionless pion energy, $f\left(y_{i}\right)=1/\left[\exp\left({y_{i}}\right)-1\right]$ is the pion distribution function, $C_{\rm a \pi}$ is the dimensionless  axion-pion coupling constant, and $\zeta(x)$ is the Riemann $\zeta$-function. The energy scale $f_{{\rm a}}=f_{{\rm PQ}}/N$ is the PQ scale, normalized by the PQ color anomaly $N$, and $m_{\pi}\simeq135~{\rm MeV}$ is the mass of a neutral pion \cite{pdg}. 
The PQ scale can be expressed in terms of the axion mass \cite{weinbergmass}:\begin{equation} f_{{\rm a}}\simeq \frac{\sqrt{r}}{1+r}\frac{f_{\pi}m_{\pi}}{m_{{\rm a}}}.\label{mapp}\end{equation} Here $r\equiv m_{{\rm u}}/m_{{\rm d}}\sim 0.56$ is the up/down quark mass ratio and $f_{\pi}\simeq 93~{\rm MeV}$ is the pion decay constant \cite{kt,pdg}. 

\begin{figure}[ht]
\includegraphics[width=3.4in]{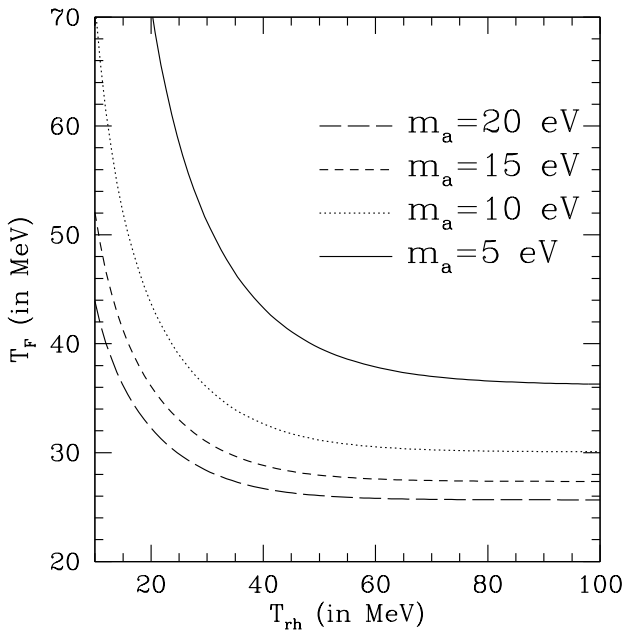}
\caption{Freeze-out temperature of the reactions $\pi^{+}+\pi^{-}\leftrightarrow \pi^{0}+a$, $\pi^{+}+\pi^{0}\leftrightarrow \pi^{+}+ a$, and $\pi^{-}+{\pi^{0}}\leftrightarrow \pi^{-}+a$, shown as a function of the reheating temperature $T_{\rm rh}$, for $4$ different axion masses. More massive axions are coupled more strongly, leading to later freeze-out than for lighter axions.}
\label{ft_dependence}
\end{figure}

Evaluating Eq.~(\ref{pionrate}) for $\Gamma$ and numerically solving Eq.~(\ref{axabev}) for $H\left( T\right)$, we estimate the axion freeze-out temperature using the condition $\Gamma\left(T_{\rm F}\right)\sim H\left(T_{\rm F}\right)$. As the reheating temperature is lowered, axions freeze out at higher temperatures due to the higher value of $H$, as shown in Fig.~\ref{ft_dependence}. As the reheating temperature is increased, the $T\propto a^{-3/8}$ epoch becomes increasingly irrelevant, and the freeze-out temperature of the axion asymptotes to its standard radiation-dominated value. Examining Eq.~(\ref{pionrate}), we see that $\Gamma \propto f_{\rm a}^{-2}\propto m_{\rm a}^{2}$, so higher-mass axions keep up with the Hubble expansion for longer and generally decouple at lower temperatures. Thus, for higher $m_{\rm a}$, a more radical change to the thermal history (even lower $T_{\rm rh}$), is needed to drive $T_{\rm F}$ to a fixed higher value. 

As axions are spin-$0$ relativistic bosons, their number density at freeze-out is $n_{\rm a}\left(T_{\rm f}\right)=\zeta{\left(3\right)}T_{\rm F}^{3}/\pi^{2}$. If we assume that axion production ceases at freeze-out, the density of axions at any subsequent time is just $n_{\rm a}\left(T_{\rm f}\right)\left[a_{\rm F}/a_{0}\right]^{3}$, where $a_{\rm F}$ is the value of the cosmological scale factor at axion freeze-out.  The reheating time scale, $t_{\rm rh}\simeq 1/\Gamma$, is much shorter than the Hubble time for $T\lsim T_{\rm rh}$, and so it is a good approximation to treat the break between the $T\propto a^{-3/8}$ and $T\propto a^{-1}$ epochs as instantaneous at $ T=T_{\rm rh}$. Doing so, we apply Eq.~(\ref{tsfac}) prior to the completion of reheating and $a\propto  T^{-1} g_{*_{{\rm S}}}^{-1/3}\left(T\right)$ afterwards, to obtain
\begin{equation}
\Omega_{\rm a}h^{2}=\frac{m_{\rm a, \rm eV}}{130}\left(\frac{10}{g_{*_{\rm S},\rm F}}\right)\gamma\left(T_{\rm rh}/T_{\rm F}\right),\nonumber
\end{equation}
\begin{equation}\gamma(\beta)\sim
\left\{\begin{array}{ll}
\beta^{5}\left(\frac{g_{*, \rm rh}}{g_{*,\rm F}}\right)^{2}\left(\frac{g_{*_{\rm S}, \rm F}}{g_{*_{\rm S}, \rm rh}}\right)&\mbox{if $\beta\ll1$,}\\ 
1&\mbox{if $\beta\gg 1$,}
\end{array}\label{newomew}\right.
\end{equation} where $m_{\rm a}$ is the axion mass in units of ${\rm eV}$. 

\begin{figure}[ht]
\includegraphics[width=3.4in]{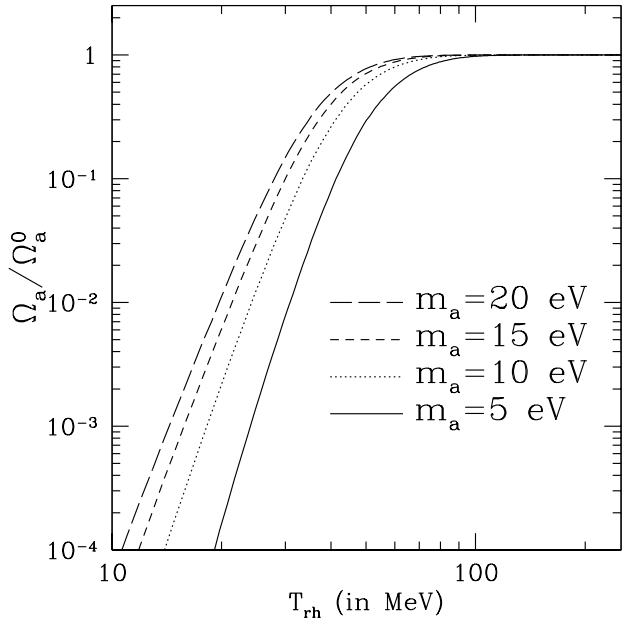}
\caption{Axion abundance $\Omega_{\rm a}$ normalized by its standard value $\Omega_{\rm a}^{0}$, shown as a function of the reheating temperature $T_{\rm rh}$. Curves are shown for $4$ different axion masses. More massive axions freeze out later. As a result, their density is less diluted by entropy production during the reheating epoch, and so they have higher relic densities.}
\label{abundance_dependence}
\end{figure}

Low reheating temperatures drive up the freeze-out temperature. When $T_{\rm rh}\lsim T_{\rm F}$, the present mass density in axions is severely suppressed, because of the sharper dependence of the scale factor $a$ on $T$ during reheating. This is a result of entropy generation. Using the numerical solution for $a\left(T\right)$ from Sec. \ref{ltrsec}, we obtain $\Omega_{\rm a}$, accounting for the smooth transition between the $T\propto a^{-3/8}$ and $T\propto a^{-1}$ regimes. In Fig.~\ref{abundance_dependence},  we show $\Omega_{\rm a}$ normalized by its standard value, $\Omega_{\rm a}^{0}$, as a function of $T_{\rm rh}$. At reheat temperatures just a factor of a few below the usual axion freeze-out temperature for a given axion mass, the axion abundance is suppressed by a factor of $10^{-4}-10^{-3}$. For $T_{\rm rh}\gg T_{\rm F}$, the axion abundance asymptotes towards its standard value. 

In the case of kination, axion freeze-out temperatures are still raised, but there is no additional entropy production. Axion abundances are given by Eq.~(\ref{thermalfreeze}), but with the higher $g_{*_{\rm S},\rm F}$ values appropriate at higher values of $T_{\rm F}$. 

For the LTR case, our results do not depend on the initial value $H_{\rm I}$ of the Hubble parameter. 
As seen in Fig.~\ref{figureeva}, changes to $H_{\rm I}$ determine the moment of the fast rise to $T_{\rm max}$, but have little influence on the expansion history for $T<T_{\rm max}$. For convenience, we choose $H_{\rm I}=1~{\rm MeV}$ for our calculations, corresponding to $T_{\rm max}\simeq 20~{\rm GeV}$. 

Our calculation is valid only if axions are produced in equilibrium by thermal pions. This requirement imposes the restriction $T_{\rm rh}\gsim 10~{\rm MeV}$. Outside this range, our assumptions break down.  For sufficiently low values of $m_{\rm a}$ and $T_{\rm rh}$, pionic cross sections lead to $T_{\rm F}\gsim 200~{\rm MeV}$, earlier than the quark-hadron phase transition. The absence of hadrons then necessitates the use of quark-axion production cross sections. 

Furthermore, for $T_{\rm rh}\lsim 10~{\rm MeV}$, pions will decay before they can come into equilibrium. In both cases, axion abundances are suppressed relative to our calculation. For axion masses saturating our upper limits and $T_{\rm rh}\gsim 10~{\rm MeV}$, we have checked that we are well within the equilibrium regime. We restrict ourselves to this range, noting that for $T_{\rm rh} \lsim 10~{\rm MeV}$, more suppressed abundances will lead to an even more dramatic relaxation to cosmological axion limits. Finally, coherent oscillations of the axion field produce a condensate that behaves as cold dark matter \cite{giudice1}, but the resulting additional abundance is negligible for $m_{\rm a}\gsim 10^{-2}~{\rm eV}$ at all values of $T_{\rm rh}$ under consideration here \cite{giudice1}.  

\section{Constraints to axions}
\label{revise}
Most constraints to the axion mass are obtained from its two-photon interaction. This interaction is parameterized by a coupling constant $g_{\rm a \gamma\gamma}$, given by
\cite{Raffelt,kt,kaplan,srednicki,kephartw,Bershady:1990sw,notinvisible,ressellt,grin,gnedin,cast,sikivie_cavity} \begin{eqnarray}
g_{\rm a \gamma \gamma}=-\frac{\alpha}{2\pi f_{\rm a}}\frac{3\xi}{4},\end{eqnarray} where $\xi$ is a model-dependent parameter and $\alpha$ is the fine-structure constant. The tightest constraint to $g_{\rm a \gamma \gamma}$,\begin{equation}
g_{a \gamma \gamma}\lsim 0.6 \times 10^{-10}~{\rm GeV}^{-1},\label{globlim}
\end{equation} comes from the helium burning lifetime of stars in star clusters \cite{Raffelt,raffelt_globular}.
The parameter $\xi$ is given by \cite{Raffelt,buckleymurayama,srednicki,kaplan}
\begin{equation}\xi =\frac{4}{3}\left[ \frac{E}{N}-\frac{2}{3}\frac{\left(4+r\right)}{\left(1+r\right)}\right]\label{defxi}.
\end{equation} Here $E$ and $N$ are weighted sums of the electric and PQ charges of fermions that couple to axions. In existing axion models, $0\leq E/N<8/3$ \cite{murayama,buckleymurayama}, while $r$ is poorly constrained and lies in the range \cite{buckleymurayama,pdg1,lqcd,kma}:
\begin{equation}
0.2\leq r\leq 0.8. \label{range}
\end{equation}

As a result, for any axion model in the range $1.93\lsim E/N<2.39$, there are experimentally allowed $r$ values for which $g_{a \gamma \gamma}$ vanishes (see Eq.~\ref{defxi}), and so constraints to axions from star clusters, helioscope, RF cavity, and telescope searches may all be lifted \cite{buckleymurayama,murayama}. Such a cancellation is fine tuned, but even for other values of $E/N$, constraints to the axion mass are loosened. 

In contrast, the hadronic couplings do not vanish for any experimentally allowed $r$ values \cite{murayama}. Axion searches based on these couplings are underway, and have already placed new upper limits to the axion mass in the ${\rm keV}$ range \cite{kekez} (and references therein).  These couplings also determine the cosmological abundance of axions, and so useful constraints may be obtained from cosmology.  

Although the hadronic coupling determines the relic abundance of axions, $\xi$ will determine the lifetime of the axion, which may have implications for cosmological constraints.  For the high axion masses allowed by our new constraints and certain values of $r$ and $E/N$, the decay $a\to \gamma\gamma$ may no longer be negligible on cosmological time scales. Such an axion would be completely unconstrained by limits to $\Omega_{a}h^{2}$ from the total matter density or the hot dark matter mass fraction. In the following calculation, we neglect axion decay. Consistent with a vanishing two-photon coupling for $E/N=2$, we use the value $r=0.50$. We have verified that our results for $\Omega_{a}h^{2}$ vary by only $5\%$ for variations in $r$ of about $20\%$, as the dependence of the axion production rate and $T_{\rm F}$ on $r$ is weak (see Eq.~\ref{pionrate}) compared with the dependence on the $e^{-m_{\pi}/{T}}$ Boltzmann factor.

\subsection{Constraints to the axion mass from $\Omega_{\rm m} h^{2}$}

In the standard cosmology, a conservative constraint is obtained by requiring that axions not exceed the total matter density of $\Omega_{\rm m}h^{2}\simeq 0.135$ \cite{spergelwmap3}, yielding the limit  $m_{\rm a}\lsim 22~{\rm eV}$, using a concordance value for the dimensionless Hubble parameter $h=0.73$. In LTR scenarios, axion abundances are highly suppressed, as shown in Eq.~(\ref{newomew}) and Fig.~\ref{abundance_dependence}. Mass constraints to thermal axions from cosmology are thus considerably relaxed. 

To obtain abundances in the LTR scenario, we apply the numerical freeze-out and abundance calculation of Section \ref{production}. Axion mass limits resulting from the requirement that axions not exceed the total dark matter density are shown by the dot-dashed hashed region in Fig.~\ref{limits}.  If $T_{\rm rh}\lsim 40~{\rm MeV}$, constraints are considerably relaxed. For example, if $T_{\rm rh}\simeq 10~{\rm MeV}$, the axion mass constraint is $m_{\rm a}\leq1.4~{\rm keV}$. When $T_{\rm rh}\gsim 95~{\rm MeV}$, we obtain $m_{\rm a}\lsim 22~{\rm eV}$, equal to the standard radiation-dominated result. As already discussed, abundances are only slightly changed in the case of kination, so $\Omega_{\rm a}h^{2}$ imposes the constraint $m_{\rm a}\lsim 26~{\rm eV}$ if $T_{\rm kin}\simeq 10~{\rm MeV}$.

\subsection{Constraints to the axion mass from CMB/LSS data}

Axions will free stream at early times, decreasing the matter power spectrum on length scales smaller than the comoving free-streaming scale, evaluated at matter-radiation equality. For sufficiently low masses,  axions will also contribute to an enhanced early ISW effect \cite{isw} in the CMB. This suppression is given by $\Delta P/P\simeq-8\Omega_{\rm a}/\Omega_{m}$ if $\Omega_{\rm a}\ll \Omega_{m}$ \cite{hanraffelt,raffeltwong,hann5} (and references therein). The matter power spectrum thus imposes a constraint to $\Omega_{\rm a}h^{2}$.

First we review the constraints imposed by these effects in a standard thermal history.  
In this case, both $\Omega_{\rm a}h^2$ and the free-streaming scale, $\lambda_{\rm fs}$, depend only on $m_{\rm a}$ in a hadronic axion model. Using Sloan Digital Sky Survey (SDSS)  measurements of the galaxy power spectrum \cite{sdss1} and Wilkinson Microwave Anisotropy Probe (WMAP) \cite{spergelwmap1} 1$^{\rm st}$-year measurements of the CMB angular power spectrum, Refs.~\cite{hanraffelt,raffeltwong,hann5} derived limits of $m_{\rm a}\lsim 1~{\rm eV}$. Axions in the mass range of interest ($m_{\rm a}\gsim 1~{\rm eV}$) are non-relativistic at photon-baryon decoupling, and so this constraint essentially comes from measurements of the galaxy power spectrum \cite{crotty}. As a result, we do not expect that an analysis including more recent WMAP results would make a substantial difference in the allowed axion parameter space. In order to lift this constraint, the relationship between $m_{\rm a}$ and $\Omega_{\rm a}h^2$ or $\lambda_{\rm fs}$ must be changed. 

If $T_{\rm F}$ is allowed to vary freely, the constraints may be relaxed.  In particular, using Eq.~(\ref{thermalfreeze}) we can see that increasing $g_{*_{\rm S},\rm F}$ (and thus $T_{\rm F}$) will decrease $\Omega_{\rm a} h^2$.  Furthermore, if the free-streaming length of the axion is less than the smallest length scale on which the linear power-spectrum may be reliably measured ($\lambda_{\rm min}\simeq 40\ h^{-1}\ {\rm Mpc}$), its effect on the matter power spectrum is not observable \cite{hanraffelt,raffeltwong,hann5}.  The comoving free-streaming length scale at matter-radiation equality\footnote{This differs from the expression in Refs.~\cite{hann5,haneutrino5} because we assume, as is the case in our parameter region of interest, that $m>T_{\rm eq}$, the temperature at matter-radiation equality. In Ref.~\cite{haneutrino5}, $m<T_{\rm eq}$ is assumed.} \cite{kt},
\begin{eqnarray}\nonumber \lambda_{\rm fs}\simeq \frac{196~{\rm Mpc}}{m_{\rm a, \rm eV}}\left(\frac{T_{\rm a}}{T_{\nu}}\right)\times \\  \left\{1+\ln{\left[0.45 m_{\rm a, \rm eV}\left(\frac{T_{\nu}}{T_{\rm a}}\right)\right]}\right\},\label{reallfs}\end{eqnarray}
is set by the ratio between the axion and neutrino temperatures, \begin{eqnarray}
\frac{T_{\rm a}}{T_{\rm \nu}}\approx(10.75/g_{*_{\rm S},\rm F})^{1/3},
\label{tconv}
\end{eqnarray} 
so that if $g_{*_{\rm S},\rm F}\gsim 87$ (i.e., if axions freeze out considerably before neutrinos), the constraint to axion masses is lifted \cite{hanraffelt}. 

\begin{figure}
\includegraphics[height=3.15in]{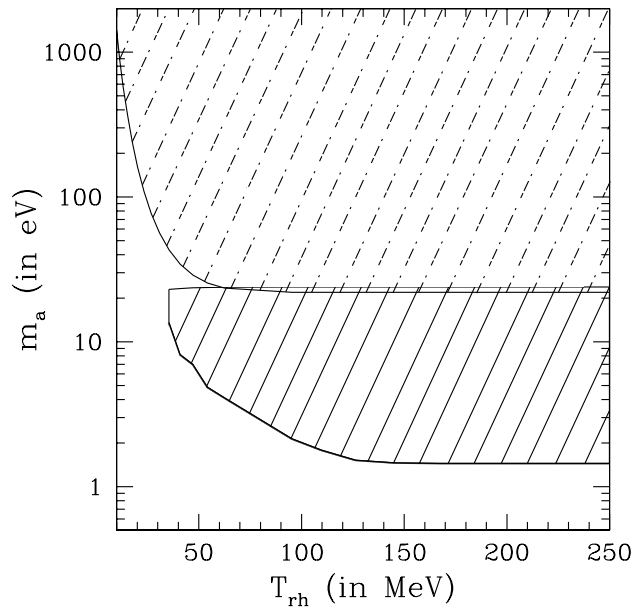}
\caption{Upper limits to the hadronic axion mass from cosmology, allowing the possibility of a low-temperature-reheating scenario. The dot-dashed hatched region shows the region excluded by the constraint $\Omega_{{\rm a}}h^{2}<0.135$ as a function of reheating temperature $T_{\rm rh}$. The solid hatched region shows the additional part of axion parameter space excluded by WMAP1/SDSS data. At low reheating temperatures, upper limits to the axion mass are loosened.  For $T_{\rm rh}\gsim 170~{\rm MeV}$, the usual constraints are recovered.}
\label{limits}
\end{figure}

In the case of a modified thermal history, the relationship between $T_{\rm F}$ and $m_{\rm a}$ acquires dependence on an additional parameter ($T_{\rm rh}$, in the case of LTR, or $T_{\rm kin}$, in the case of kination) thus allowing us to loosen the constraints. At a series of values of $g_{*_{\rm S},\rm F}$, Refs.~\cite{hanraffelt,raffeltwong,hann5} determine the maximum values of $\Omega_{\rm a}h^{2}$ consistent with WMAP measurements of CMB  power spectra and SDSS measurements of the galaxy power spectrum. We begin by mapping these contours, from Fig.~5b in Ref.~\cite{hanraffelt} (which do not include constraints from the Lyman-$\alpha$ forest), into the $\left(\Omega_{a}h^{2},\lambda_{\rm fs}\right)$ plane. For a fixed $m_{\rm a}$ or $\Omega_{\rm a}$, $\lambda_{\rm fs}$ scales monotonically with $g_{*_{\rm S},\rm F}$, and thus serves as a proxy for $g_{*_{\rm S},\rm F}$.

In the domain $10~{\rm MeV}\leq T_{\rm rh}\leq 250~{\rm MeV}$ and $0.01~{\rm eV}\lsim m_{\rm a}\lsim 10~{\rm keV}$, we calculate $\Omega_{\rm a}\left(T_{\rm rh},m_{\rm a}\right)h^{2}$ for hadronic axions in LTR, using the full numerical solution described in Sec. \ref{production}. We also calculate $\lambda_{\rm fs}\left(T_{\rm rh},m_{\rm a}\right)$.
Since axions freeze out while relativistic, their energy will redshift as $E\propto a^{-1}$. They will have temperature  $T_{\rm a}=T_{\rm F}a_{\rm F}/a$. Meanwhile, the temperature of the coupled radiation redshifts as $T \propto a^{-3/8}$ until radiation domination begins. 
Thus entropy generation modifies the relationship between the axion and neutrino temperatures to
\begin{eqnarray}
\frac{T_{a}}{T_{\nu}}\simeq \left(\frac{11}{4}\right)^{1/3}\left(\frac{T_{\rm rh}}{T_{\rm F}}\right)^{5/3}
\left(\frac{g_{*,\rm rh}^{2}g_{*_{\rm S},0}}{g_{*,\rm F}^{2}g_{*_{\rm S},\rm rh}}\right)^{1/3},
\end{eqnarray} 
if $T_{\rm F}>T_{\rm rh}$. To obtain all of our constraints we use the more precise scaling accounting for the smooth transition between the $T\propto a^{-3/8}$ and $T\propto a^{-1}$ regimes. The dominant change to the free-streaming length comes from the modified axion temperature, while the modified expansion rate itself induces negligible fractional changes of order $T_{\rm NR}/T_{\rm rh}$, where $T_{\rm NR}$ is the cosmic temperature at which the axion goes non-relativistic.

For each pair $(T_{\rm rh},m_{\rm a})$, we calculate $\Omega_{\rm a}h^{2}$ and $\lambda_{\rm fs}$ to trace out the region forbidden with $95\%$ confidence. When $\Omega_{\rm a}\left( T_{\rm rh},m_{\rm a}\right)h^{2}>0.014$, outside the domain of Ref.~\cite{hann5}, we extrapolate, assuming that the $95\%$ contour asymptotes to a line of constant axion free-streaming wavelength $\lambda_{\rm fs}=40 ~h^{-1}~{\rm Mpc}$. Such a trend is noted in Ref.~\cite{hann5}, and at the maximum value of $\Omega_{\rm a}h^{2}$ of the contour obtained from Ref.~\cite{hanraffelt}, the maximum allowed free-streaming length is consistent with our assumed asymptote.

We obtain the upper limit to the axion mass as a function of $T_{\rm rh}$, shown in Fig.~\ref{limits}.  Existing LSS/CMB constraints are severely relaxed in the LTR scenario, and lifted completely for $T_{\rm rh}\lsim 35~{\rm MeV}$. For $T_{\rm rh}\lsim35~{\rm MeV}$, $\lambda_{\rm fs}<40~h^{-1}~{\rm Mpc}$ for $m_{\rm a}\gsim1~{\rm eV}$, and so the axion mass is unconstrained. It will still be subject to phase-space constraints if it saturates the bound $\Omega_{\rm a}h^{2}\lsim 0.135$, and is to compose all the dark matter in galactic halos \cite{gt,mad1,mad2}. At high reheating temperatures, the constraint from LSS/CMB data ($\Omega_{\rm a}h^{2}\lsim0.006$) supercedes the constraint $\Omega_{\rm a}h^2\lsim\Omega_{\rm m}h^{2}$.

The narrow allowed region between the LSS/CMB and total matter density constraints in Fig.~\ref{limits} ($45~{\rm MeV}\lsim T_{\rm rh}\lsim 55~{\rm MeV}$) may be simply understood. Axions in this narrow window are cold and massive enough to evade large-scale structure constraints (i.e., $\lambda_{\rm fs}< \lambda_{\rm min}$), and dilute enough to evade constraints from the total matter density. We note that the CMB/LSS limits asymptote to their standard value of $m_{\rm a} \lsim 1.4~{\rm eV}$ for $T_{\rm rh}\gsim170~{\rm MeV}$. 

\begin{figure}[t]
\includegraphics[width=3.2in]{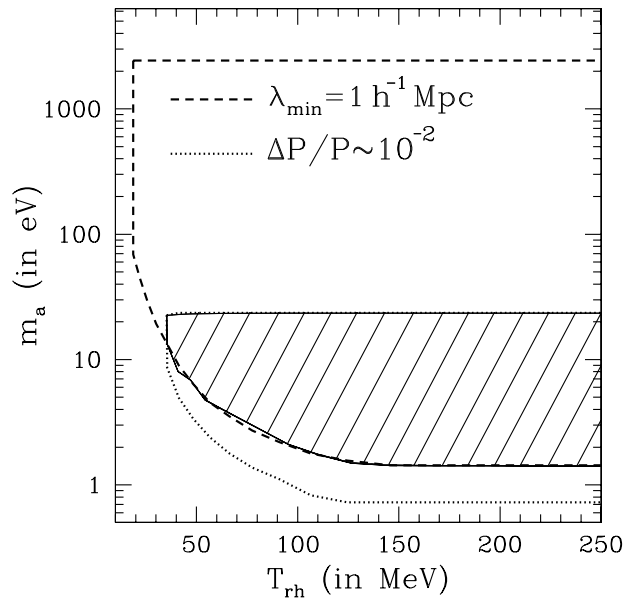}
\caption{Estimated improvement in the accessible axion parameter space from including more precise measurements of the matter power spectrum (region bounded by the dotted line), corresponding to LSST \cite{lsstintro,lsstzhan}, or from measurements of clustering on smaller length scales, corresponding to Lyman-$\alpha$ forest measurements (region bounded by the dashed line) \cite{viel}. 
The hatched region indicates the parameter space excluded by WMAP1/SDSS measurements.}
\label{lim_fut}
\end{figure}

Future instruments, such as the Large Synoptic Survey Telescope (LSST), will measure the matter power-spectrum with unprecedented precision ($\Delta P/P \sim 0.01)$ \cite{lsstzhan,lsstintro}. This order of magnitude improvement over past work \cite{sdss2,sdsslrg} will improve the constraint to $\Omega_{a}h^{2}$ by an order of magnitude, resulting in the improved sensitivity to axion masses and reheating temperatures shown by the dotted line in Fig.~\ref{lim_fut}. To estimate possible constraints to axions from LSST measurements of the power spectrum, we recalculated our limits using the approximate scaling $\Delta P/P\simeq -8\Omega_{\rm a}/\Omega_{\rm m}$, assuming $\Delta P/P \sim 10^{-2}$ for $\lambda>40~h^{-1}~{\rm Mpc}$.

We also estimate the possible improvement offered by including information on smaller scales ($\lambda_{\rm min}\sim 1~h^{-1}~{\rm Mpc}$), as may be obtained from measurements of the Lyman-$\alpha$ flux power spectrum \cite{viel}, also shown in Fig.~\ref{lim_fut}.  We include this effect by replacing $\lambda_{\rm min}$ with this lower minimum length scale.  This is indicated by the dashed line in Fig.~\ref{lim_fut}.  We can see that more massive axions are probed because of information on smaller length scales, as are lower reheating temperatures. 

In the case of kination, a much less severe relaxation of limits to axions is obtained. As there is no entropy generation in the kination case, the abundance and temperature of the axion are still given by Eqs.~(\ref{thermalfreeze}) and (\ref{tconv}), with the value of $g_{*_{\rm S},\rm F}$ appropriate at the new freeze-out temperature. In the range of parameter space explored, $10~{\rm MeV}\lsim T_{\rm F}\lsim 100~{\rm MeV}$, and so the variation in $g_{*_{\rm S},\rm F}$ as a result of kination is $\sim 60\%$. For $T_{\rm kin}\simeq 10~{\rm MeV}$, the new allowed regions are $m_{\rm a}\lsim 3.2 ~{\rm eV}$ and $17~{\rm eV}\lsim m_{\rm a}\lsim 26~{\rm eV}$. These conclusions apply to any non-entropy-generating scenario in which $H\propto T^{3}$ at some early epoch, and not only to kination \cite{kamion}. If $T_{\rm rh}\gsim 110~{\rm MeV}$, standard results are recovered.

\section{Axions as relativistic degrees of freedom at early times}
\begin{figure}[t]
\includegraphics[width=2.8in]{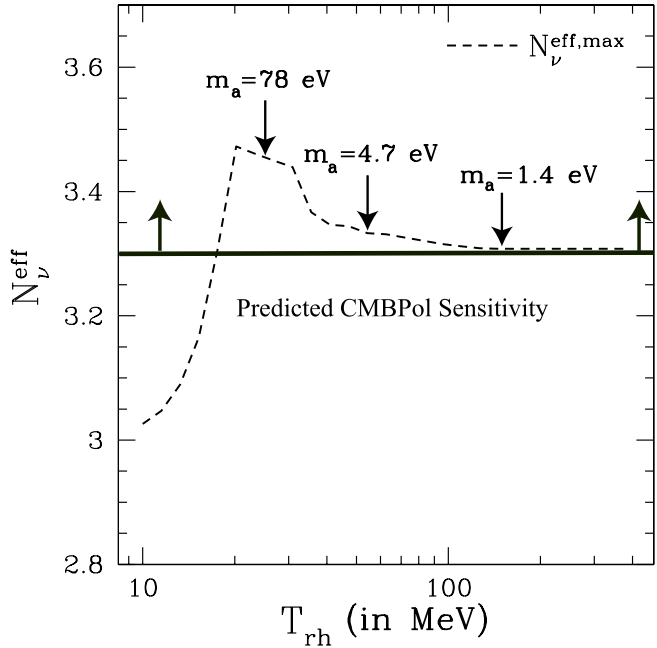}
\caption{Total effective neutrino number $N_{\nu}^{\rm eff,\rm max}$ for axions with masses saturating the tightest bound on axion masses from Fig.~\ref{limits}. The requisite higher temperatures lead to earlier axion freeze out and lower $N_{\nu}^{\rm eff,\rm max}$. The thick black line indicates the anticipated sensitivity of CMBPol \cite{kapling} to $N_{\nu}^{\rm eff}$ through the primordial helium abundance.}
\label{neff_dependence}
\end{figure}

Future limits to axions in the standard radiation-dominated and LTR thermal histories may follow from constraints to their contribution to the energy density in relativistic particles at $T\sim 1~{\rm MeV}$.
Axions are relativistic spin-$0$ bosons, and so $\rho_{\rm a}\simeq\left(\pi^{2}/30\right)T_{\rm F}^{4}\left(a_{\rm F}/a\right)^{4}=\left(\pi^{2}/30\right)T_{\rm F}^{4}\left(a_{\rm F}/a_{\rm rh}\right)^{4}\left(a_{\rm rh}/a\right)^{4}$ \cite{kt}. 
We can express the total relativistic energy density in terms of an effective neutrino number
\begin{eqnarray}
N_{\nu}^{\rm eff}\equiv\left(\frac{\rho_{\rm a}+\rho_{\nu}}{\rho_{\gamma}}\right)\left(\frac{8}{7}\right)\left(\frac{11}{4}\right)^{4/3},~~\rho_{\gamma}=\frac{\pi^{2}}{15}T^{4}\nonumber, \\
\rho_{\nu}=\frac{7}{8}\left(\frac{4}{11}\right)^{4/3} \times 3\times\left(\frac{\pi^{2}T^{4}}{15}\right).
\end{eqnarray} 
Treating the transition between the $T\propto a^{-3/8}$ and $T\propto a^{-1}$ epoch as instantaneous, we solve for the photon and axion temperatures, and then obtain
\begin{equation}
\begin{array}{r}
N_{\nu}^{\rm eff}=3+\frac{4}{7}\left(\frac{43}{4}\right)^{4/3} \Psi \left(T_{\rm F}/T_{\rm rh}\right),\nonumber
\end{array}\nonumber\end{equation}
\begin{equation}
\Psi \left(y\right)\sim\left\{\begin{array}{ll}
\left[g_{*_{\rm S},\rm rh}y^{5}\left(\frac{g_{*,\rm F}}{g_{*,\rm rh}}\right)^{2}-1\right]^{-4/3}
&\mbox{if $y\gg1$,}\\
\left[g_{*_{\rm S},\rm F}-1\right]^{-4/3}&\mbox{if $y\ll 1$}.
\end{array}\right.\label{neffans}
\end{equation} 
For sufficiently high masses, the axionic contribution saturates to $\delta N_{\nu}^{\rm eff}=4/7$ at high reheating temperatures \cite{changchoi}. In Fig.~\ref{neff_dependence}, we show $N_{\nu}^{\rm eff,\rm max}\left(T_{\rm rh}\right)$, the effective neutrino number evaluated at the axion mass which saturates the LSS/CMB bounds, for $T_{\rm rh}\gsim 35~{\rm MeV}$, or saturates the constraint $\Omega_{\rm a}h^{2}\sim 0.135$ for lower $T_{\rm rh}$. The behavior of the curve may be readily understood.  As can be seen from Fig.~\ref{limits}, as we increase $T_{\rm rh}$, the maximum allowed $m_{\rm a}$ decreases. For $T_{\rm rh} \lsim 20\ \mathrm{MeV}$, even though the maximum allowed $m_{\rm a}$ is large (which corresponds to a lower $T_{\rm F}$, since $\Gamma \propto m_{\rm a}^2$), the amount of entropy production between $T_{\rm F}$ and $T_{\rm rh}$ leads to a small axionic contribution to $N_{\rm eff}$.  As $T_{\rm rh}$ increases, the interval between freeze-out and reheating decreases. This lessens the impact of entropy generation, and leads to the rise in $N_{\rm eff}$. Finally, for $T_{\rm rh}\gsim 20~{\rm MeV}$, the impact of entropy generation is nearly negligible, and $N_{\rm eff}$ falls as the maximum allowed value of $m_{\rm a}$ decreases, as in the standard case (due to earlier freeze-out). 

A comparison between the abundance of $^{4}$He and the predicted abundance from BBN places constraints to the radiative content of the Universe at $T \sim 1$ MeV \cite{cardell}; this can be stated as a constraint to $N_{\rm \nu}^{\rm eff}$. At early times, axions will contribute to the total relativistic energy density (through $N_{\rm \nu}^{\rm eff}$), and thus constraints to $^{4}$He abundances can be turned into constraints on $m_{\rm a}$ and $T_{\rm rh}$, as shown in Fig.~\ref{neff_dependence}.  

In terms of the baryon-number density $n_b$, we write the primordial $^4$He abundance as $Y_{\rm p} \equiv 4 n_{{\rm He}}/n_b$.  In order to translate measurements of $Y_{\rm p}$ to constraints on $m_{\rm a}$ and $T_{\rm rh}$ we use the scaling relation \cite{steigmanreview}
\begin{equation}\Delta N^{\rm eff}_{\nu} = \frac{43}{7}\left\{\left(6.25\Delta Y_{p}+1\right)^{2}-1\right\}.\end{equation}
Constraints to $N_{\nu}^{\rm eff}$ from direct measurements of $Y_p$, including a determination of $n_b$ from CMB observations, lead to the 68\% confidence level upper limit of $N_{\nu}^{\rm eff} \leq 3.85$ \cite{cyburt,ichikawa2,ichikawa4}.  From Fig.~\ref{neff_dependence} and Eq.~(\ref{neffans}), we see that this bound cannot constrain $m_{\rm a}$ or $T_{\rm rh}$.  If future measurements reduce systematic errors, constraints to $T_{\rm rh}$ will be obtained for the lighter-mass axions.

Constraints to $m_{\rm a}$ and $T_{\rm rh}$ may also follow from indirect CMB measurements of $Y_p$.   The presence of $^4$He affects CMB anisotropies by changing the ionization history of the universe \cite{trotta}. The Planck satellite is expected to reach $\Delta Y_{\rm p} =0.013$, yielding a sensitivity of $N_{\nu}^{\rm eff} \leq 4.04$, while CMBPol (a proposed future CMB polarization experiment) is expected to approach $\Delta Y_p = 0.0039$, leading to the sensitivity limit $N_{\nu}^{\rm eff} \leq 3.30$ \cite{trotta,kapling,ichikawa4,hut}. As shown in Fig.~\ref{neff_dependence}, for $T_{\rm rh}\gsim 15~{\rm MeV}$, such measurements of $Y_{\rm p}$ may impose more stringent limits on the axion mass. Also, if axions with mass in the $~\rm eV$ range are directly detected, $Y_{p}$ might impose a surprising \textit{upper} limit to $T_{\rm rh}$ \cite{cast,admx}.

\section{Conclusions}
The lack of direct evidence for radiation domination at temperatures hotter than $1~{\rm MeV}$ has
motivated the introduction of kination, low-temperature reheating, and other scenarios for an altered pre-BBN expansion history. In the case of kination, the change in axion abundances and thus cosmological constraints is modest. Low-temperature reheating will suppress the abundance of thermally-produced hadronic axions, once the reheating temperature $T_{\rm rh}\sim 50~{\rm MeV}$. This is rather intuitive once we recall that the axion freeze-out temperature in a radiation dominated cosmology is $\sim 50~{\rm MeV}$. If the reheating temperature crosses this threshold, axion densities are severely reduced by dramatic entropy production during reheating. 

Total density, large-scale structure, and microwave background constraints to axions are all severely loosened as a result, possibly pushing the the axion mass window to very high values; for $T_{\rm rh}\simeq 10~{\rm MeV}$, the new constraint is $m_{\rm a}<1.4~{\rm keV}$. For $T_{\rm rh}\gsim 170~{\rm MeV}$, standard radiation dominated results are recovered. The inclusion of information on smaller scales will probe higher axion masses and lower reheat temperatures. More precise measurements of the matter power spectrum on all scales will probe lower axion masses. Kination also relaxes constraints to axions, though much less markedly. Future probes of primordial helium abundance will either lead to further constraints on axion properties, or, if axions are directly detected, provide a new view into the thermal history of the universe during the epoch $10~{\rm MeV}\lsim T\lsim 170~{\rm MeV}$.
\begin{acknowledgments}
D.G. was supported by the Gordon and Betty Moore Foundation and acknowledges helpful discussions with Mark Wise, Stefano Profumo, and Sean Tulin. T.L.S. and M.K. were supported by DoE DE-FG03-92-ER40701, NASA NNG05GF69G, and the Gordon and Betty Moore Foundation. 
\end{acknowledgments}

   \end{document}